**Chapter Eight**

# Local field topology behind light localization and metamaterial topological transitions

*Jonathan Tong, Alvin Mercedes, Gang Chen\*, and Svetlana V. Boriskina[#]*
*Department of Mechanical Engineering, Massachusetts Institute of Technology Cambridge, MA, USA*
*[\*]gchen2@mit.edu, [#]sborisk@mit.edu*

We revisit the mechanisms governing the sub-wavelength spatial localization of light in surface plasmon polariton (SPP) modes by investigating both local and global features in optical powerflow at SPP frequencies. Close inspection of the instantaneous Poynting vector reveals formation of optical vortices – localized areas of cyclic powerflow – at the metal-dielectric interface. As a result, optical energy circulates through a subwavelength-thick 'conveyor belt' between the metal and dielectric where it creates a high density of optical states (DOS), tight optical energy localization, and low group velocity associated with SPP waves. The formation of bonding and anti-bonding SPP modes in metal-dielectric-metal waveguides can also be conveniently explained in terms of different spatial arrangements of localized powerflow vortices between two metal interfaces. Finally, we investigate the underlying mechanisms of global topological transitions in metamaterials composed of multiple metal and dielectric films, i.e., transitions of their iso-frequency surfaces from ellipsoids to hyperboloids, which are not accompanied by the breaking of lattice symmetry. Our analysis reveals that such *global topological transitions* are governed by the dynamic local re-arrangement of *local topological features* of the optical interference field, such as vortices and saddle points, which reconfigures global optical powerflow within the metamaterial. These new insights into plasmonic light localization and DOS manipulation not only help to explain the well-known properties of SPP waves but also provide useful guidelines for the design of plasmonic components and materials for a variety of practical applications.

## 8.1 INTRODUCTION

Tailored light interactions with metal surfaces and nanostructures can generate coherent collective oscillations of photons and free electrons – known as surface plasmons [1-3]. These hybrid collective states can be supported both by planar metal-dielectric interfaces in the form of surface plasmon polariton (SPP) waves and by metal particles and nanostructures in the form of localized surface plasmon (LSP) modes. Unique physical characteristics of plasmonic modes include extreme spatial field localization, high density of optical states (DOS), and low group velocity. These features open up new opportunities for nanoscale trapping and manipulation of light with applications in sensing [4-7], spectroscopy [8-13], on-chip communications [14, 15], and solar energy harvesting [16-24].

It has recently been revealed that some of the unique features associated with plasmonic effects on nanoparticles and particle clusters can be explained by the unusual pathways of nanoscale optical powerflow in the immediate vicinity of the metal nanostructures [25-29]. The local optical powerflow at each point in space is uniquely defined by the presence of local topological features in the phase of the optical interference field close to this point. Local topological features include phase singularities – points or lines in space where the field intensity is zero due to destructive interference. At these phase singularities, the phase field is undefined. Other types of local topological features are stationary phase nodes – points or lines in space where the phase gradient vanishes [27, 30-33]. The stationary phase nodes include local

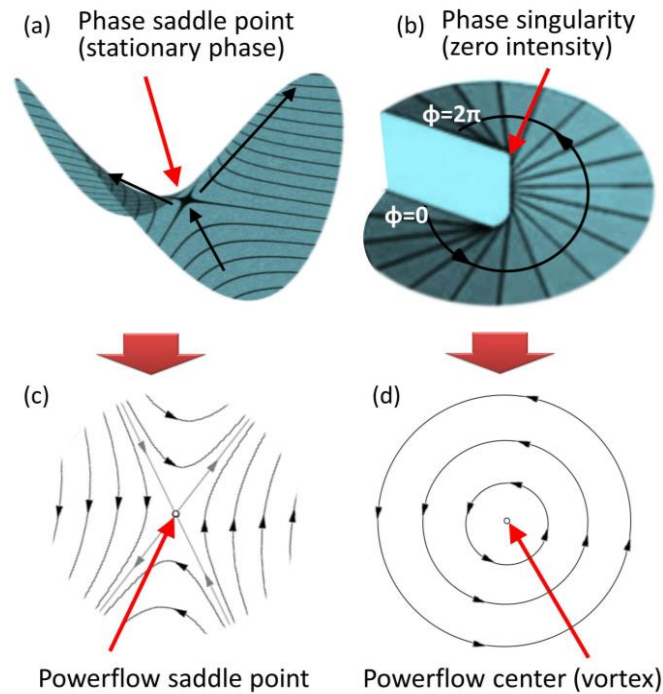

Figure 8.1. Local topological features in the optical phase field and powerflow. (a) Phase saddle node – a stationary point (or line) where the phase gradient vanishes. (b) Optical phase singularity – a point (or line) of destructive interference where the field intensity vanishes and the phase is undefined (i.e., all values of phase from 0 to $2\pi$ co-exist). (c) The optical powerflow saddle point corresponding to the phase saddle point in (a). (d) The optical powerflow vortex corresponding to the phase singularity in (b).

extrema (i.e., phase maxima and minima) as well as phase saddle nodes as shown in Fig. 8.1a.

Because the optical power always flows in the direction of the phase change, the phase maxima (minima) give rise to the powerflow sinks (sources) while phase saddle points give rise to saddle nodes in the powerflow (Fig. 8.1c). Likewise, optical phase singularities (illustrated in Fig. 8.1b) are accompanied by the circulation of optical powerflow, resulting in the formation of so-called optical vortices (or centers of power flow) as shown in Fig. 8.1d. Both saddle points and vortices are classified as local topological features because they are characterized by conserved quantities such as the topological charge [30, 34, 35]. The topological charge is the net phase change in a closed loop enclosing the local feature (quantized in units of 2π), which is positive if the phase increases in a right-handed sense. Overall, local topological features '*constitute a 'skeleton' on which the phase and intensity structure hangs*' [32].

The role of local topological features in the modification of nanoscale powerflow causing the well-known phenomenon of enhanced light absorption by small metal nanoparticles – with the absorption cross-section larger than the geometric cross-section – has been revealed by Craig Bohren in 1983. He demonstrated the reversal of optical powerflow in the shadow region behind the particle [36] illuminated by an electromagnetic plane wave at a frequency corresponding to the particle dipole surface plasmon mode. This flow reversal follows the local optical phase landscape, which is shaped by the presence of local topological features such as powerflow saddle points (Fig. 8.2a) above and inside the particle [27, 37-39]. In all the panels of Fig. 8.2, the large orange arrows indicate the direction of the incident wave, while the little arrows or streamlines illustrate the direction of the powerflow. The powerflow intensity at each point is defined by either the arrow length or the streamline density. The corresponding intensity distribution of the electric field is plotted in the background. Two decades later, it was found that local topological features in the near-field region of metal nanoparticles can form not only at the frequency of their local surface plasmon resonance (LSP) but also under off-resonance illumination by light as shown in Fig. 8.2b [39]. These features include powerflow saddle points as well as nanoscale optical vortices (Fig. 8.2b) [39]. The complex near-field phase landscape around isolated nanoparticles can be manipulated by tuning their sizes, shapes, and materials characteristics. Continuous change of these parameters results in nucleation, spatial drift, and annihilation of local topological features [37, 40].

It was later discovered that new plasmonic effects emerge due to tailored coupling of nanoscale powerflows generated around individual particles. In particular, the local powerflow picture driven by the formation and coupling of nanoscale optical vortices helped to explain the extreme nano-focusing of light in snowmen-like plasmonic nanolenses composed of three neighboring nanospheres of progressively smaller size [26, 27, 41]. This is illustrated in Fig. 8.2c, which shows how the circulating powerflows formed on each sphere merge and recirculate the optical energy through the narrow inter-particle gaps, resulting in the build-up of intensity in the gaps. As illustrated in Fig. 8.2d, optical vortices formed in the vicinity of nanoscale plasmonic antennas due to light scattering from neighboring antennas can lead to enhanced light absorption in or enhanced light scattering from the nanoantennas [42].

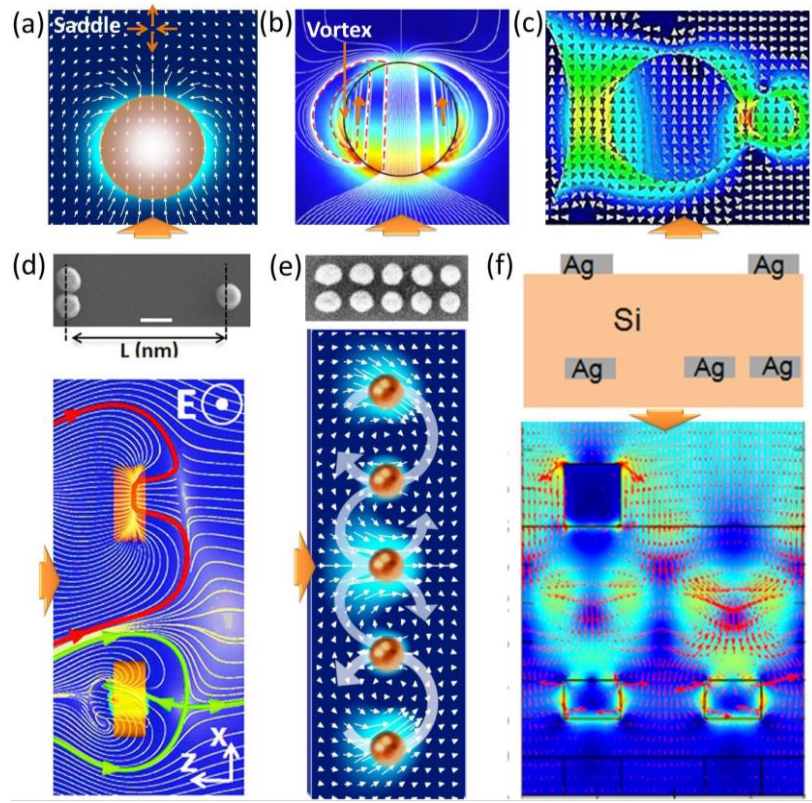

Figure 8.2. Local-topology-driven powerflow around metal nanoparticles. (a) Enhanced light absorption by a nanoparticle owing to the powerflow saddle point in its shadow (adapted with permission from [27] ©RSC). (b) Counter-rotating optical vortices on a nanosphere (adapted with permission from [33] ©OSA). (c) Coupled optical vortices in the plasmonic nanolens (adapted with permission from [27] ©RSC). (d) Optical vortices induced on plasmonic antennas by light scattered from another antenna (reproduced with permission from [36] © WILEY). (e) Coupled optical vortices 'pinned' to a linear chain of nanoparticle dimers form a linear transmission-like sequence (adapted with permission from [25] ©ASC). (f) Vortices formed in the Silicon slab due to light scattering from embedded metal nanoparticles (adapted with permission from [38] ©OSA).

However, proper engineering of optical powerflow around plasmonic nanostructures can enable suppression of both optical scattering and material absorption. In particular, it has been shown that to reduce dissipative losses in metals, the structures need to be designed to 'pin' optical vortices that recirculate optical energy outside of the metallic particles (Fig. 8.2e) [25-29]. This approach enables generation of narrow-band resonant features in the optical spectra of plasmonic nanostructures. It can also be used to achieve enhanced absorption in the host material (e.g. semiconductor) rather than in metal, as shown in Fig. 8.2f. These results therefore pave the way for vortex-pinning nanostructures to be used to enhance absorption in thin-film photovoltaic cells [25-27, 29, 43, 44].

The results presented in Fig. 8.2 clearly demonstrate that formation of local topological features and the resulting recirculation of the optical power through metal nanostructures are behind many interesting plasmonic effects such as extreme nano-focusing of light and electric field intensity enhancement. In this chapter, we demonstrate that localized topological features play a much larger role in various plasmonic effects than it has been recognized to date. We start with the simplest SPP wave on a metal-dielectric interface and then extend the analysis to metal-insulator-metal waveguides and to so-called hyperbolic metamaterials [45]. Close inspection of the local field phase profiles and optical powerflow in the vicinity of the metal surfaces reveals formation of localized areas of circulating powerflow that recycles optical energy between metal and dielectric volumes. Such recirculation translates into the tight light localization of the SPP mode field on the surface, the large in-plane wavevector of SPP waves, and the resulting slowing of the wave propagation along the metal-dielectric interfaces.

In the following sections, we will discuss how the insights into the spatial structure of the localized topological features in the near field of plasmonic components and materials help to better understand their properties, to predict and exploit new plasmonic effects, and to design next generation plasmonic devices with improved performance.

## 8.2 BACK TO BASICS: SURFACE PLASMON POLARITON

The most studied and written about plasmonic effect is the excitation of surface plasmon polariton waves propagating along metal-dielectric interfaces. SPP modes are TM polarized surface waves, which are characterized by large wavevectors parallel to the interface and low group velocities. As a result, they create strong electric field and high local density of optical states (LDOS) within the sub-wavelength-thick layer adjacent to the interface. These waves can only exist on interfaces between materials having dielectric permittivity values of opposite signs. Dielectric permittivities of materials with a high density of free charge carriers – such as metals and highly doped semiconductors – are defined by the Drude model, $\varepsilon_m(\omega) = \varepsilon_\infty - \omega_p^2/(\omega^2 + i\gamma\omega)$, where $\omega_p^2$ is the plasma frequency, $e_\yen$ is the high-frequency permittivity limit, and $g$ is the electron collision frequency. The real part of the Drude permittivity becomes negative in the frequency range below the plasma frequency of the material. This makes possible SPP propagation along their interfaces with other materials.

An eigensolution of the electromagnetic boundary problem for the Maxwell equations on such an interface (shown in Fig. 8.3a) that describes the SPP mode has the following form [15]:

$z > 0$:
$$H_y(z) = Ae^{i\beta x}e^{-k_2 z}$$
$$E_x(z) = iA\frac{k_2}{\omega\varepsilon_0\varepsilon_2}e^{i\beta x}e^{-k_2 z} \quad (8.1)$$
$$E_z(z) = -A\frac{\beta}{\omega\varepsilon_0\varepsilon_2}e^{i\beta x}e^{-k_2 z}$$

and

$z < 0$:
$$H_y(z) = Ae^{i\beta x}e^{k_1 z}$$
$$E_x(z) = iA\frac{k_1}{\omega\varepsilon_0\varepsilon_1}e^{i\beta x}e^{k_1 z} \quad (8.2)$$
$$E_z(z) = -A\frac{\beta}{\omega\varepsilon_0\varepsilon_1}e^{i\beta x}e^{k_1 z}$$

Here, $A$ is an arbitrary amplitude of the magnetic field, $\omega$ is the angular frequency, $\varepsilon_0$ is the vacuum permittivity, $\varepsilon_i$ is the relative permittivity of the $i$-th medium, $\beta$ is the component of the wave vector parallel to the interface, and $k_i$ is the component of the wavevector in the *i*-th medium normal to the interface. Upon applying continuity boundary conditions to field expressions (8.1) and (8.2), a dispersion relation can be obtained as follows:

$$\beta = k_0 \sqrt{\frac{\varepsilon_1 \varepsilon_2}{\varepsilon_1 + \varepsilon_2}} \, , \tag{8.3}$$

where $k_0$ is the vacuum wave vector. The solution to equation (8.3) is plotted in Fig. 8.3b, and the branch corresponding to the SPP mode has a familiar flat dispersion form, reaching to infinite momentum values in the absence of dissipative losses in metal. Of course some level of losses is inevitable in real materials; however, to make the following argument simple and straightforward without the loss of generality, we are going to consider an idealized case of lossless materials (i.e., $g = 0$). In this case, the longitudinal component of the SPP wave vector that is a solution to dispersion equation (8.3) is purely real, while the normal component is purely imaginary. For all subsequent calculations in this section, the permittivities of the dielectric and metal are chosen to be silicon dioxide and silver respectively [46, 47]. For Ag, we only used the data for the real part of the permittivity and neglected the dissipative term. However, as will be shown in the next section, the analysis and conclusions are valid for the general case of materials with dissipative losses.

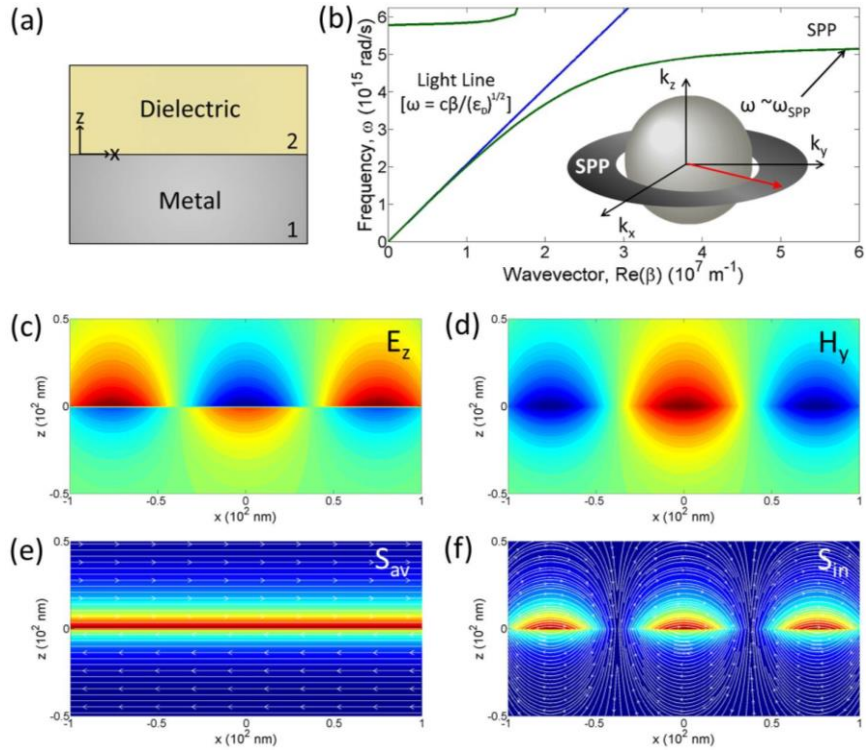

Figure 8.3. Circulating powerflow behind tight light localization and low group velocity in Surface Plasmon Polariton waves. (a) Schematic of the planar metal-dielectric interface and the coordinate system used in the analysis. (b) The dispersion characteristics of the waves supported by the structure in (a). The inset shows an iso-frequency surface of the allowed photon momenta above the interface at the frequency close to the plasma frequency in the plasmonic material. (c,d) The electric (c) and magnetic field components of the SPP wave. (e) The time-averaged optical powerflow around the material interface. (f) The corresponding instantaneous powerflow at *t*=0. All the field distributions were calculated at $w$=4.974x10$^{15}$ rad/s.

The momentum-space iso-frequency surface for the dielectric material just above the interface around the SPP frequency is shown in the inset to Fig. 8.3b [29]. This surface contains all the allowed k-vectors, and is a combination of a sphere corresponding to the propagating waves and a ring corresponding to the high-momentum SPP branch of the dispersion equation (8.3). In the absence of dissipative losses, the ring extends to infinity, which is in stark contrast with the finite-size spherical iso-frequency surface of regular materials. Even if dissipative losses are present, the number of allowed photon momenta at the interface is dramatically increased, resulting in the high local density of optical states (LDOS) and thus in the high electromagnetic energy density of the SPP mode. However, a high LDOS and a high intensity of both electric and magnetic fields (Figs. 8.3c,d) are only observed in close vicinity to the material interface. Away from the interface, the LDOS and intensity drop off exponentially.

To get deeper insight into the mechanisms that squeeze the optical energy of the SPP wave to the region just around the interface, we study the optical powerflow at the SPP frequency. The direction and intensity of the time-averaged optical powerflow at each point of space can be characterized by the Poynting vector, which is calculated as follows:

$$\langle S \rangle = \frac{1}{2} Re(E \times H^*). \tag{8.4}$$

In the simple geometry of Fig. 8.3a, we can derive an analytical form of the time-averaged Poynting vector for each medium. Substitution of (8.1) and (8.2) into (8.4) yields the following expressions for the time-averaged powerflow above and below the interface:

$$z > 0: \langle S \rangle = \hat{x}\frac{A^2}{2}\frac{\beta}{\omega\varepsilon_0\varepsilon_2}e^{-2k_2 z} \tag{8.5}$$

$$z < 0: \langle S \rangle = \hat{x}\frac{A^2}{2}\frac{\beta}{\omega\varepsilon_0\varepsilon_1}e^{2k_1 z} \tag{8.6}$$

Comparison of (8.5) and (8.6) immediately shows that the power flow parallel to the interface reverses direction abruptly when crossing from one medium to the other. The plot of the time-averaged powerflow shown in Fig. 8.3e visualizes the above observation. The time-averaged Poynting vector component in the vertical direction has a purely imaginary value, which is a signature of a reactive powerflow. In electromagnetic circuits, reactive power is the portion of power associated with the stored energy that returns to the source in each cycle and transfers no net energy to the load [48]. Reactive power is always a factor in alternating current circuits such as electrical grids, where energy recycling through storage elements (i.e., inductors and capacitors) causes periodic reversals in the direction of energy flow. Although the reactive powerflow does not deliver any useful energy to the load, it assists in maintaining proper voltages across the power system. Reactive powerflow is manifested in measurable dissipative losses due to periodic energy recycling through the grid and sudden disruptions in the reactive powerflow pattern can cause a voltage drop along the line [49]. Likewise, the *reactive powerflow away from the metal-dielectric interface* associated with the *SPP propagation along the interface* does not contribute to the net energy transfer. The power is temporarily stored in the form of magnetic and electric fields.

As the SPP mode is an eigensolution of the Maxwell equations rather than a wave generated by either a localized source or a plane wave, an insight into its energy storage mechanism can be gained by plotting the instantaneous Poynting vector distribution [50]. The instantaneous optical powerflow at any given moment in time can be calculated as follows:

$$S = \frac{1}{2}Re(E \times H^*) + \frac{1}{2}Re(E \times He^{i\omega t}). \tag{8.7}$$

It can be seen that the time-averaged powerflow expression (8.4) is represented by the first term in (8.7), and the second term drops out due to its sinusoidal behavior when performing the averaging procedure. The instantaneous Poynting vector of the SPP wave is plotted in Fig. 8.3e and features repeating areas of circulating optical powerflow centered on the metal-dielectric interface. These data lead us to the conclusion that electromagnetic energy recycling through local optical vortices on the interface is behind tight field localization, high energy density, and reduced group velocity of the SPP waves. The new look at the old problem provided by revealing the circulating powerflow also offers a new take on the meaning of the large in-plane *k*-vector photon states that *only exist very close to the interface* supporting SPP waves. The circulating instantaneous SPP optical powerflow is characterized by the photon *angular momentum*, which has a conserved value at every point in space [51]. In turn, the value of the *linear momentum* that is tangential to the circulating powerflow scales inversely with the distance to the powerflow center (i.e., with the distance to the metal-dielectric interface). The tangential momentum in the direction perpendicular to the interface is canceled out by the time averaging, leaving only the in-plane tangential momentum, which, in turn, drops off away from the interface.

## 8.3 SPP COUPLING VIA SHARED CIRCULATING POWERFLOW

Another interesting class of plasmonic waveguiding platforms is a metal-insulator-metal (MIM) structure, which guides optical energy along a narrow slot between two metal interfaces. In such plasmon slot waveguides optical mode volumes can be reduced to sub-wavelength scales while suffering low decay even for frequencies far from the plasmon resonance [52]. A schematic of the MIM plasmon slot waveguide is shown in Figs. 8.4a and 8.5a, and the dispersion characteristics are plotted in Figs. 8.4b and 8.5b. The dispersion characteristics are obtained by solving the matrix equation with the appropriate boundary conditions on both metal-dielectric interfaces:

$$\text{Odd mode}: \tanh(k_1 a) = -\frac{k_2 \varepsilon_1}{k_1 \varepsilon_2}. \tag{8.8}$$

$$\text{Even mode}: \tanh(k_1 a) = -\frac{k_1 \varepsilon_2}{k_2 \varepsilon_1}. \tag{8.9}$$

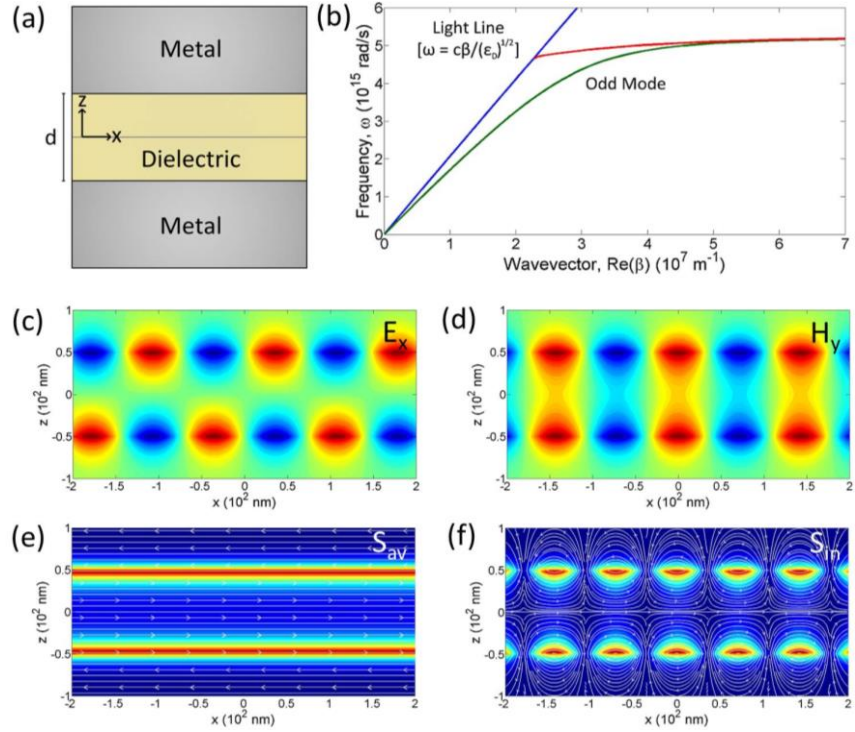

Figure 8.4. Shared circulating powerflow of the odd SPP mode in a MIM plasmon slot waveguide. (a) Schematic of the MIM waveguide and the coordinate system. (b) The dispersion characteristics of the SPP modes supported by the structure in (a). The green curve is the dispersion branch corresponding to the odd mode. (c,d) The electric (c) and magnetic field components of the odd SPP mode. (e) The time-averaged optical powerflow in the MIM waveguide. (f) The corresponding instantaneous powerflow at $t=0$. All field distributions were calculated at $w=4.974 \times 10^{15}$ rad/s.

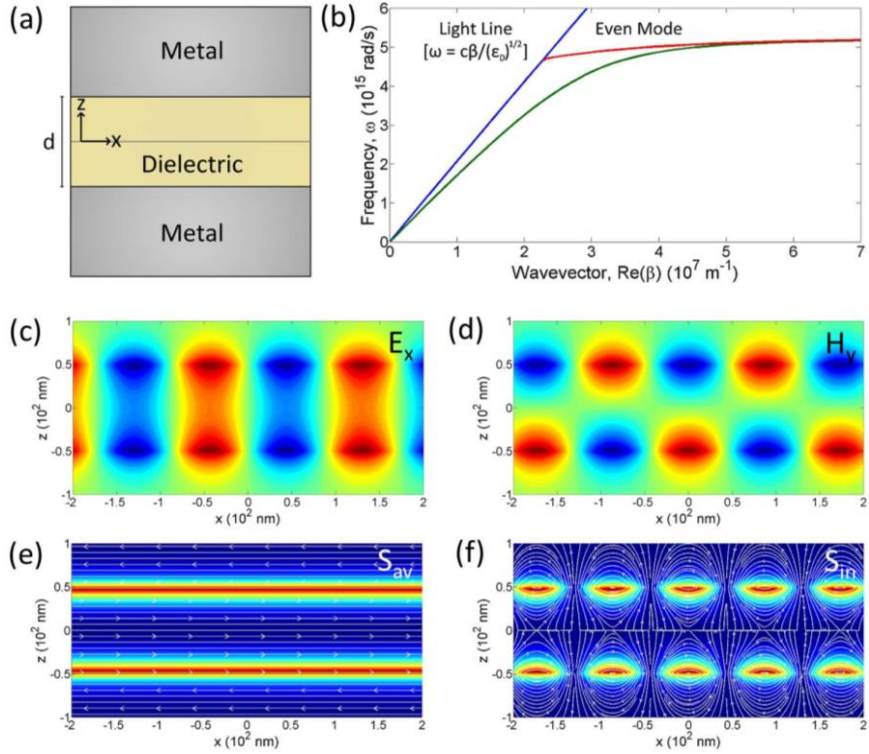

Figure 8.5. Shared circulating powerflow of the even SPP mode in a MIM plasmon slot waveguide. (a) Schematic of the MIM waveguide and the coordinate system. (b) The dispersion characteristics of the SPP modes supported by the structure in (a). The red curve is the dispersion branch corresponding to the even mode. (c,d) The electric (c) and magnetic field components of the even SPP mode. (e) The time-averaged optical powerflow in the MIM waveguide. (f) The corresponding instantaneous powerflow at *t*=0. All field distributions were calculated at $w=4.974 \times 10^{15}$ rad/s.

Dispersion relations (8.8) and (8.9) correspond to the modes with tangential electric field distributions that are either symmetric (even mode) or anti-symmetric (odd mode) with respect to the waveguide axis. The near-field patterns of the electric and magnetic fields of the odd SPP mode are plotted in Figs. 8.4c and 8.4d, respectfully, and the time-averaged optical powerflow is shown in Fig. 8.4e. Similar to the case of a single metal-dielectric interface, the time-averaged power inside and outside the dielectric slot waveguide flows in opposite directions. The recycling of energy between the backward and forward flow channels is achieved via coupling of circulating powerflows around each interface, which is revealed in the instantaneous Poynting vector distribution plotted in Fig. 8.4f. Figure 8.5 illustrates the same characteristics of the even coupled-SPP mode of the plasmon slot waveguide, which corresponds to the higher-energy branch in the waveguide dispersion characteristics (see Fig. 8.5b). The differences in the coupling mechanism underlying formation of even and odd modes are illustrated in Figs. 8.4f and 8.5f. The instantaneous Poynting vector field of the odd coupled-SPP mode is characterized by merging of the circulating powerflows formed on each material interface (Fig. 8.4f). In contrast, the circulating powerflows of the even SPP mode collide at the slot waveguide center (Fig. 8.5f), which results in the blue-shift of the corresponding dispersion curve (Fig. 8.5b).

## 8.4 HYPERBOLIC METAMATERIALS: GLOBAL FIELD TOPOLOGY DEFINED BY LOCAL TOPOLOGICAL FEATURES

Even more interesting optical effects can be engineered via near-field optical coupling between SPP waves formed on multiple stacked M-I interfaces. The resulting anisotropic nanostructured metal-dielectric material is schematically shown in Fig. 8.6a. Mutual electromagnetic coupling of SPP modes across multiple M-I interfaces results in the formation of several SPP branches in the dispersion characteristics of such anisotropic structures (shown in Fig. 8.6b). Here and in the following figures the thicknesses of the Ag and $TiO_2$ layers are 9nm and 22nm, respectively, and the dissipative losses in Ag are fully accounted for. The presence of multiple high-k branches seen in Fig. 8.6b increases the bandwidth of the plasmonic-enhanced high-DOS spectral region over that of the SPP on a single interface [29].

Accurate solutions for the dispersion of the multilayered metamaterial plotted in Fig. 8.6b were obtained by using the analytical transfer matrix method. However, in the limit of an infinite number of ultra-thin layers, the multilayered metamaterials shown in Fig. 8.6a can be described within an effective index model by a uniaxial effective dielectric tensor $\hat{\varepsilon} = diag[\varepsilon_{xx}, \varepsilon_{yy}, \varepsilon_{zz}]$ [45, 53-58]. If $\varepsilon_{xx} = \varepsilon_{yy} = \varepsilon_\parallel$ and $\varepsilon_\parallel \cdot \varepsilon_{zz} < 0$, the anisotropic metamaterial dispersion relation

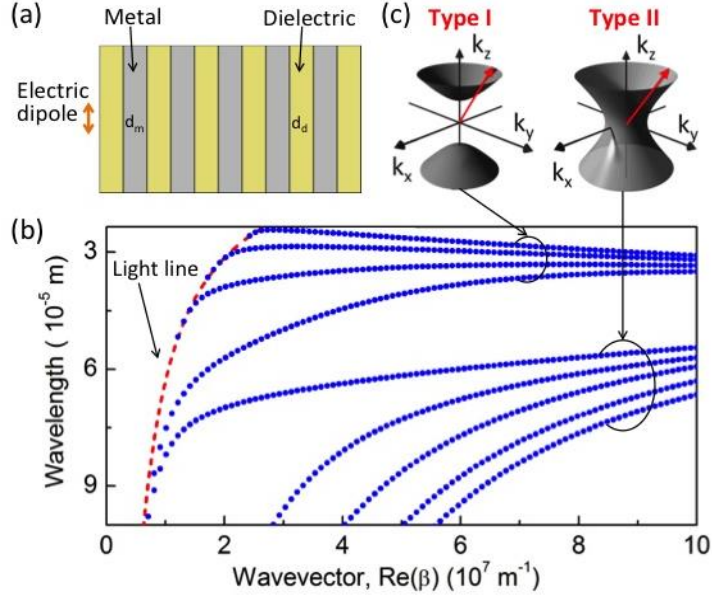

Figure 8.6. Metal-dielectric hyperbolic metamaterials: SPP modes dispersion and iso-frequency surfaces. (a) Schematic of a multilayered metamaterial (metal layers are yellow-colored). (b) Modal dispersion characteristics of the metamaterial. (c) Momentum-space representation of the dispersion relations (iso-frequency surfaces) in two frequency ranges where the metamaterial undergoes a topological transition from an ordinary material to HMM, with allowed wavevectors shown as red arrows.

$\omega^2/c^2 = (k_x^2 + k_y^2)/\varepsilon_{zz} + k_z^2/\varepsilon_\parallel$ transforms from having a closed shape, i.e., ellipsoid, to an open one, i.e., hyperboloid as shown in Fig. 8.6c [45, 53-58]. This topological transition from an ordinary to a so-called hyperbolic photonic metamaterial (HMM) is an analog of the Lifshitz topological transition in superconductors [59] when their Fermi surfaces undergo transformation under the influence of external factors such as pressure or magnetic fields. In the same manner that the Lifshitz transition leads to dramatic changes in the electron transport in metals [60, 61], the topological transition of the photonic metamaterial has a dramatic effect on its DOS and photon transport characteristics. This enables development of novel devices with enhanced optical properties including super-resolution imaging, optical cloaking, and enhanced radiative heat transfer [45, 53-58]. The spectral region that includes the high-energy branches of the dispersion corresponds to the metamaterial transformation into a type I hyperbolic regime. The type I HMM is characterized by a single negative component of the dielectric tensor, which is perpendicular to the interface. A type II HMM (corresponding to the spectral region overlapping low-energy dispersion branches)

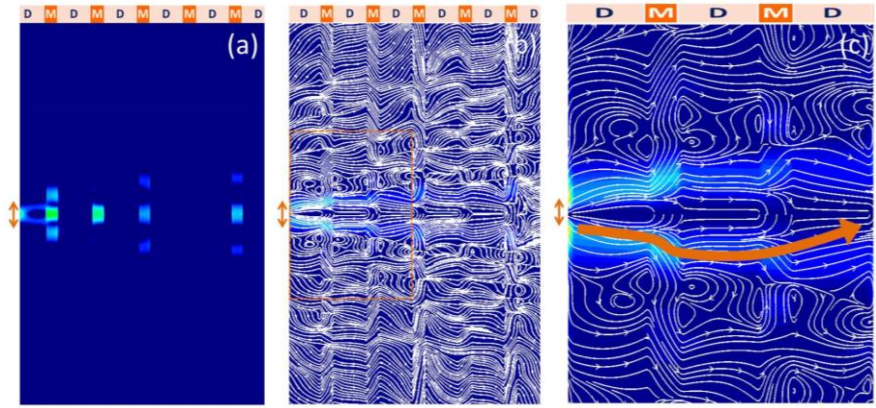

Figure 8.7. Dipole radiation into the type I hyperbolic metamaterial. (a) Electric field intensity distribution inside a 10-period-thick Ag/TiO$_2$ multilayer structure generated by a classical electric dipole shown as the orange arrow. (b) Optical powerflow from the dipole source through the metamaterial. (c) A magnified view of the region in (b) inside the orange rectangle. In (b) and (c), the Poynting vector streamlines (white) indicate the local direction of the powerflow, and the background color map shows the time-averaged Poynting vector intensity distribution. The orange arrows in (c) highlight the global powerflow direction. In all the panels, the excitation wavelength is 350nm.

features two negative in-plane components of the dielectric tensor.

In the following, we will reveal that the global topological transition of the multilayered metamaterial into the hyperbolic regime is driven by the collective dynamics of local topological features in the electromagnetic field such as nucleation, migration, and annihilation of nanoscale optical vortices. It has been already shown both theoretically and experimentally that the high DOS in the hyperbolic metamaterial strongly modifies radiative rates of quantum emitters such as quantum dots positioned either inside or close to the HMM [53, 55, 56]. To reveal the local electromagnetic field topology underlying this process, we consider radiation of a classical electric dipole located in air just outside the HMM slab (as shown Figs. 8.6-8.9). The dipole is separated from the metamaterial slab by a 10nm-thick $TiO_2$ spacer layer, which is a typical configuration in the experiment that helps to avoid quenching of quantum emitters such as quantum dots via non-radiative energy transfer to the metamaterial slab [53, 55, 56].

Figures 8.7 and 8.8 demonstrate how the metamaterial in the type I and type II hyperbolic regimes modifies radiation and transport of photons emitted by a dipole source. The availability of high-k states in the HMM provides additional channels for the dipole to radiatively decay and increases the electromagnetic energy density within the metamaterial. As a result, the dipole source emission pattern becomes strongly asymmetrical with most of the energy being channeled through the metamaterial. Furthermore, as revealed by Fig. 8.7, for the metamaterial in the type I hyperbolic regime, energy transport through the material is very directional. Most of the optical power flows through a narrow channel across the metamaterial without experiencing significant lateral spread within the metamaterial slab. A detailed picture of the local topological features that drive the directional powerflow across the metamaterial is shown in Figs. 8.7b and 8.7c. It should be noted that in Figs. 8.7-8.9 we plot *the time-averaged powerflow* defined by (8.4) rather than the instantaneous powerflow. Multiple areas of coupled vortex powerflow are clearly visible and their spatial arrangement favors an overall directional power flow by preventing lateral energy spread. The local-topology-driven directional energy transport across the HMM slab also results in the directional emission of the energy by the HMM surface, which can be used for the design of directional light sources [62].

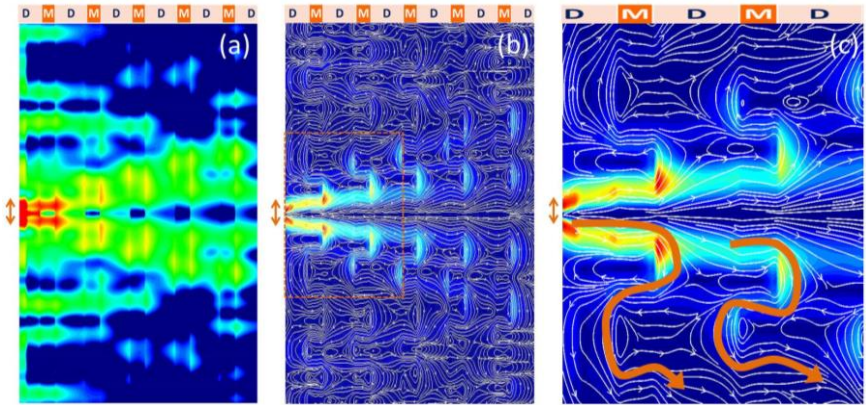

Figure 8.8. Dipole radiation into the type II hyperbolic metamaterial. (a) Electric field intensity distribution inside a 10-period-thick Ag/$TiO_2$ multilayer structure generated by a classical electric dipole shown as the orange arrow. (b) Optical powerflow from the dipole source through the metamaterial. (c) A magnified view of the region in (b) inside the orange rectangle. In (b) and (c), the Poynting vector streamlines (white) indicate the local direction of the powerflow, and the background color map shows the time-averaged Poynting vector intensity distribution. The orange arrows in (c) highlight the global powerflow direction. In all the panels, the excitation wavelength is 650nm.

Likewise, the high DOS within the metamaterial in the type II hyperbolic regime results in emission from the dipole source that is predominantly into the material. However, the powerflow driven by the emitted photons is markedly different from the type I HMM. In contrast, the optical powerflow through the type II HMM features significant lateral spread, as shown in Fig. 8.8. The local topology of the electromagnetic interference field in the metamaterial slab plotted in Figs. 8.8b and 8.8c reveals the mechanism driving the lateral energy spread. Once again, the formation of multiple coupled counter-rotating nanoscale optical vortices drives the global powerflow by recirculating the energy around local circulation points. The arrangement of the vortices in this case is different from that in the type I HMM, resulting in the markedly different overall powerflow pattern.

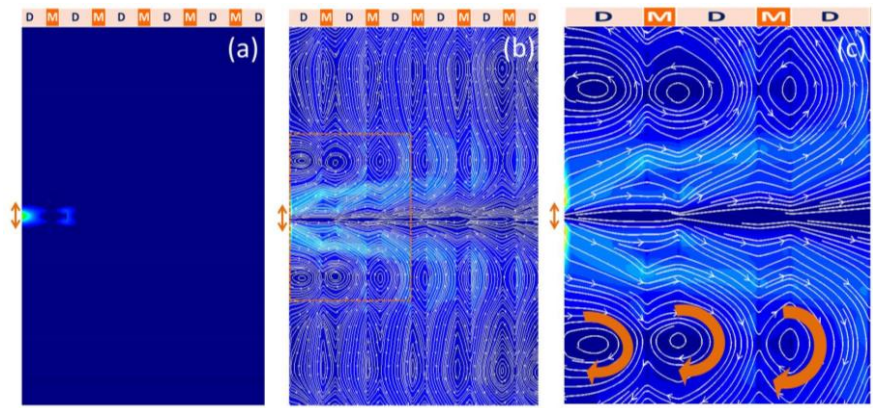

Figure 8.9. Dipole radiation into the multilayer stack outside of the hyperbolic regime. (a) Electric field intensity distribution inside a 10-period-thick Ag/TiO$_2$ multilayer structure generated by a classical electric dipole shown as the orange arrow). (b) Optical powerflow from the dipole source through the multilayer. (c) A magnified view of the region in (b) inside the orange rectangle. In (b) and (c), the time-averaged Poynting vector streamlines (white) indicate the local direction of the powerflow, and the background color map shows the Poynting vector intensity distribution. The orange arrows in (c) highlight the global powerflow direction. In all the panels, the excitation wavelength is 500nm.

Finally, Fig. 8.9 illustrates the dramatic changes in the powerflow through the metamaterial in the photon energy range where the material in not in the hyperbolic regime. In this case, the metamaterial DOS is significantly lower than that in either type of HMM. This reduces the dipole radiative rate resulting in the low energy density within the material as shown in Fig. 8.9a. The time-averaged Poynting vector field within the material still features areas of circulating powerflow as shown in Figs. 8.9b and 8.9c. However, in this case the vortices once again are spatially rearranged and form a global energy recirculation network that inhibits powerflow across the metamaterial slab. Overall, we can conclude that the re-arrangement of the local field topology features is behind the global topological transitions in hyperbolic metamaterials.

## 8.5  CONCLUSIONS AND OUTLOOK

We have demonstrated that the formation of optical vortices – localized areas of circulating optical powerflow – is a hidden mechanism behind many unique characteristics of surface plasmon polariton modes on metal-dielectric interfaces. These include tight energy localization in the vicinity of the M-I interface and 'structural slow light' characterized by the reduced group velocity of SPP waves. Furthermore, a detailed understanding of the SPP powerflow characteristics helps to explain the existence of photon states with high linear lateral momentum only in close proximity to the interface. This is a manifestation of the angular momentum of photons recycled through coupled optical vortices formed on the interface. Furthermore, our analysis reveals that the formation and dynamical reconfiguring of connected networks of coupled optical vortices underlies the global topological transitions of artificial M-I materials into the hyperbolic regime.

The above results provide further support for a rational strategy we recently developed for the design of photonic components with novel functionalities [26, 27]. This bottom-up design strategy is based on the accurate positioning of local phase singularities and connecting them into coupled networks with the aim of tailoring the global spatial structure of the interference field. Understanding the origins and exploiting wave effects associated with phase singularities have proven to be of high importance in many branches of physics, including hydrodynamics, acoustics, quantum physics, and singular optics [63, 64]. To date, the most significant advances in optics driven by controllable formation of optical vortices have been related to the generation of propagating light beams and fiber modes carrying orbital angular momentum. This research has far-reaching applications in optical trapping and manipulation and offers a way to increase the data transmission rates via angular momentum multiplexing mechanisms [51, 65, 66]. By revealing the role of optical vortices in the unique characteristics of surface plasmon waves we not only explain well-known plasmonic effects but also offer a new bottom-up approach to design plasmonic nanostructures and metamaterials with tailored energy transport characteristics. This can pave the way to new applications of plasmonic materials in optical communications [27, 67, 68], energy harvesting from solar and terrestrial heat sources [29], radiative heat transfer [29], imaging [69], and sensing [25, 26, 28].

## Acknowledgments

This work has been supported by the US Department of Energy DOE-BES Grant No. DE-FG02-02ER45977 (for near-field energy transport) and by the US Department of Energy under the SunShot grant No. 6924527 (for alternative ways of manipulating photon trapping and recycling).

8.5 Conclusions and outlook | 21